\documentclass[prl,twocolumn,showpacs,preprintnumbers,amsmath,amssymb]{revtex4}

\usepackage{graphicx}
\usepackage{dcolumn}
\usepackage{bm}

\begin{document}

\title{Measured Instantaneous Viscous Boundary Layer in Turbulent Rayleigh-B\'{e}nard Convection}

\author{Quan Zhou$^{1,2}$ and Ke-Qing Xia$^1$}

\address{$^1$Department of Physics, The Chinese University of Hong Kong, Shatin, Hong Kong, China \\
$^2$Shanghai Institute of Applied Mathematics and Mechanics, Shanghai University, Shanghai 200072, China}
\date{\today}

\begin{abstract}
We report measurements of the instantaneous viscous boundary layer
(BL) thickness $\delta_v(t)$ in turbulent Rayleigh-B\'{e}nard
convection. It is found that $\delta_v(t)$ obtained from the
measured instantaneous two-dimensional velocity field exhibits
intermittent fluctuations. For small values, $\delta_v(t)$ obeys a
lognormal distribution, whereas for large values the distribution
of $\delta_v(t)$ exhibits an exponential tail. The variation of
$\delta_v(t)$ with time is found to be driven by the fluctuations
of the large-scale mean flow velocity and the local horizontal
velocities close to the plate can be used as an instant measure of
this variation. It is further found that in the present parameter range of the experiment the mean velocity profile
measured in the laboratory frame can be brought into
coincidence with the theoretical Prandtl-Blasius laminar BL profile, if it
is resampled relative to the time-dependent frame of
$\delta_v(t)$.

\end{abstract}


Accepted for publication in \textbf{Phys. Rev. Lett.}

\maketitle

An important issue in the study of fluid dynamics is to determine
the velocity within a very thin layer in the neighborhood of the
plates and walls, i.e. the viscous boundary layer (BL) \cite{schlichting04}. The
classical Prandtl-Blasius BL, derived more than 100 years ago for
flows over a flat plate, remains one of the few BLs with exact
theoretical profile based on the equations of motion. For
turbulent boundary layers, the well-known, empirical, logarithmic
``law of wall" can be used to describe the shape of the velocity
profile near the boundary. However, there are situations in which
the BL is neither fully turbulent nor strictly laminar, but
fluctuates with time. How to characterize such BLs quantitatively
has long been a challenge. Here, we present a method that
expresses the measured boundary layer quantities in a
time-dependent frame that fluctuates with the boundary layer
thickness. We use the turbulent Rayleigh-B\'{e}nard (RB)
convection as the test case and show that the mean velocity
profile measured in the laboratory frame can be brought into
coincidence with the theoretical Prandtl-Blasius laminar BL profile, if it
is resampled relative to the time-dependent frame of BL thickness.
More generally, the concept and the algorithm presented in this
work may serve as an example on how to disentangle the mixed dynamics
in quantities from local measurements.

The Rayleigh-B\'{e}nard system has become a paradigm for the study
of general convection phenomena occurring in nature
 \cite{agl_lx}. There already exist a number of studies of the viscous BL
in RB convection \cite{belmonte, xia, sano97pre, qiuxia,
adrain02etfs, lam02pre, verzicco03jfm, thess07prl, sun08jfm},
however, an important issue remains unresolved. Scaling wise and
in a time averaged sense, it has been found that the measured BL
properties are the same as those of the Prandtl-Blasius-type laminar BL
\cite{qiuxia, sun08jfm, agl_lx}. Whereas, the time-averaged velocity
profile is found to differ from both the Prandtl-Blasius-type laminar BL
and the turbulent logarithmic BL \cite{thess07prl}, especially for
the region around the thermal BL where thermal plumes are
generated. We note that, as the viscous BL is produced
and stabilized by the viscous shear of the large-scale mean flow,
the fluctuations of the large-scale velocity would cause the
viscous BL to fluctuate as well, which would in turn cause a fixed
measurement point to be sometimes inside and sometimes outside the
BL \cite{sun08jfm}. In this respect, the time-averaged properties
are not sufficient to reveal the dynamic nature of the BL. As
fluctuating BLs exist in various flow systems, a more effective
analysis method is thus highly desirable to unlock the intricate
flow dynamics in the vicinity of plates, which is the objective of
the present work.

Two water-filled rectangular cells were used in the experiments,
the details of which  have been described elsewhere
\cite{xia03pre}. The length, width, and height, in unit of cm, of
the small cell are $25\times7\times24$ and those of the large cell
are $81\times20\times76$. The experiment was conducted at fixed Prandtl number Pr $=4.3$. The small-cell measurements covered the range
$10^9\lesssim$ Ra $\lesssim2\times10^{10}$ of the Rayleigh number
Ra and the large-cell measurements were made in the regime
$2\times10^{10}\lesssim$ Ra $\lesssim7\times10^{11}$. The technique of
particle image velocimetry (PIV) was used to measure the velocity
fields above the center of the bottom plate and within the
vertical circulation plane of the large-scale mean flow. The
measuring region for the large (small) cell experiments has an
area of $17.2\times21.5$ ($11.1\times13.8$) mm$^2$, corresponding
to $63\times79$ velocity vectors, with a spatial resolution 0.27
(0.17) mm. The large (small) cell measurement for each Ra lasted
4.42 (3.79) hours in which a total of 35000 (30000) vector maps
were acquired at a sampling rate $\sim$2.2 Hz. The time-averaged
statistical properties and various scaling behaviors of the
viscous BL measured in the small cell have been reported
previously \cite{sun08jfm}. Here we make new analysis to the data
and present the results alongside with those from the new
measurements in the large cell.

\begin{figure}
\begin{center}
\resizebox{1\columnwidth}{!}{%
  \includegraphics{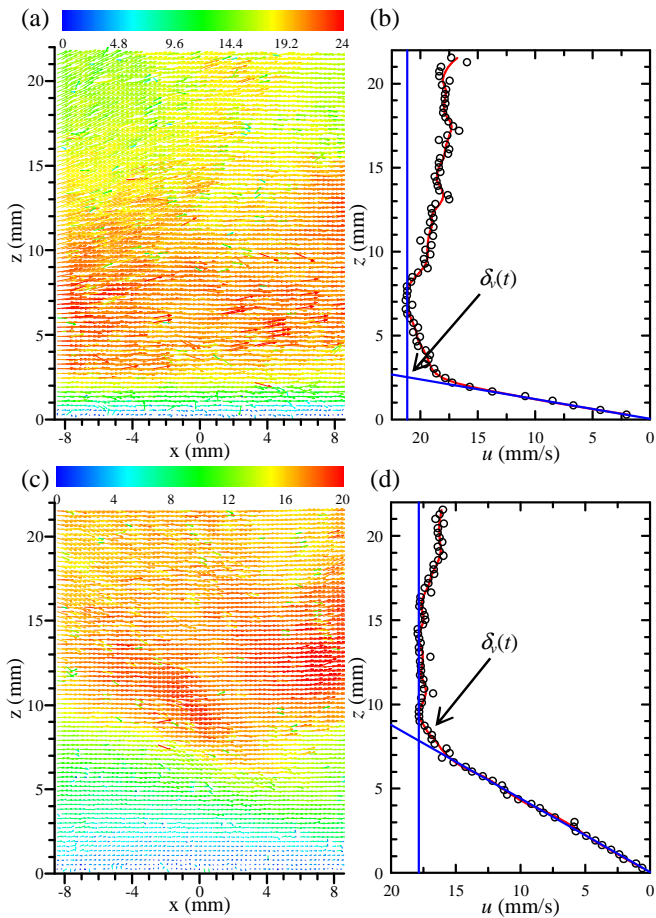}}
\caption{\label{fig:fig1} (color online) (a, c) Two examples of the
instantaneous velocity fields measured near the center of the
bottom plate (Ra $=1.9\times10^{11}$). (b, d) The horizontal
velocity profiles $u(z,t)$ corresponding to those in (a) and (c),
but averaged over $-2$ mm $\leq x \leq 2$ mm. The magnitude of the
velocity in (a,c) is coded in both color scale and the length of
the arrows in unit of mm/s. The solid lines in (b,d) illustrate
how $\delta_v(t)$ is obtained.}
\end{center}
\end{figure}

Figures \ref{fig:fig1}(a) and \ref{fig:fig1}(c) show two examples
of the instantaneous velocity fields. It is seen that the velocity
fields near the plate show strong fluctuations and larger velocity
magnitude yields thinner viscous BL. To obtain more quantitative
information, we calculate this fluctuating BL thickness from the
instantaneous velocity field using the following procedure. To
reduce data scatter, we first coarse grain the measured velocity
field by averaging it along the $x$-direction over a range of 2 mm
for the small cell and 4 mm for the large cell. This yields the
horizontal velocity profile $u(x, z, t)$ at time $t$. As the
profile shows only weak $x$-dependence, we will hereafter present
and discuss results measured at the center of the plate, i.e.
$u(z, t)\equiv u(0, z, t)$ [circles in Figs. \ref{fig:fig1}(b) and
\ref{fig:fig1}(d)]. Next, to increase the statistical accuracy,
each profile is smoothed by using the locally weighted scatterplot
smoothing method \cite{loess}, denoted as $u_s(z,t)$ [solid curves in Figs.
\ref{fig:fig1}(b) and \ref{fig:fig1}(d)]. A linear fitting is then
made to $u_s(z,t)$ close to the plate. Denote
the maximum velocity of $u_s(z,t)$ as $u_m(t)$, the instantaneous
viscous BL thickness $\delta_v(t)$ is then obtained as the
distance from the plate at which the extrapolation of the linear
fitting of $u_s(z,t)$ crosses $u_m(t)$. The uncertainty of
$\delta_v(t)$, primarily due to uncertainties of $u_m(t)$ and the
linear fitting to $u_s(z,t)$ close to the plate, is estimated to
be about $3\%$. We note that the results are robust and
insensitive to the algorithm used to smooth the profile.

\begin{figure}
\begin{center}
\resizebox{1\columnwidth}{!}{%
  \includegraphics{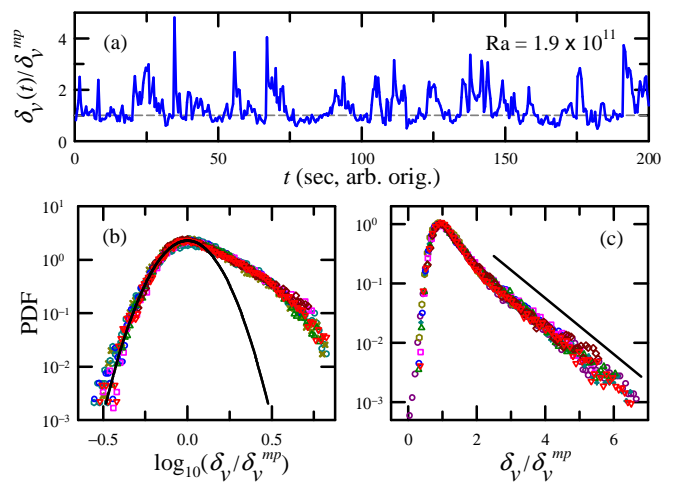}}
\caption{\label{fig:fig2} (color online) (a) A sample time trace of instantaneous
viscous BL thickness $\delta_v(t)$, normalized by its
most-probable value $\delta_v^{mp}$. (b) PDFs of
$\log_{10}(\delta_v/\delta_v^{mp})$  and (c) those of
$\delta_v/\delta_v^{mp}$ measured at various Ra varying from
$1.25\times10^9$ to $6.4\times10^{11}$. The solid curve in (b)
marks the lognormal distribution for reference.}
\end{center}
\end{figure}

\begin{figure*}
\begin{center}
\resizebox{2\columnwidth}{!}{%
  \includegraphics{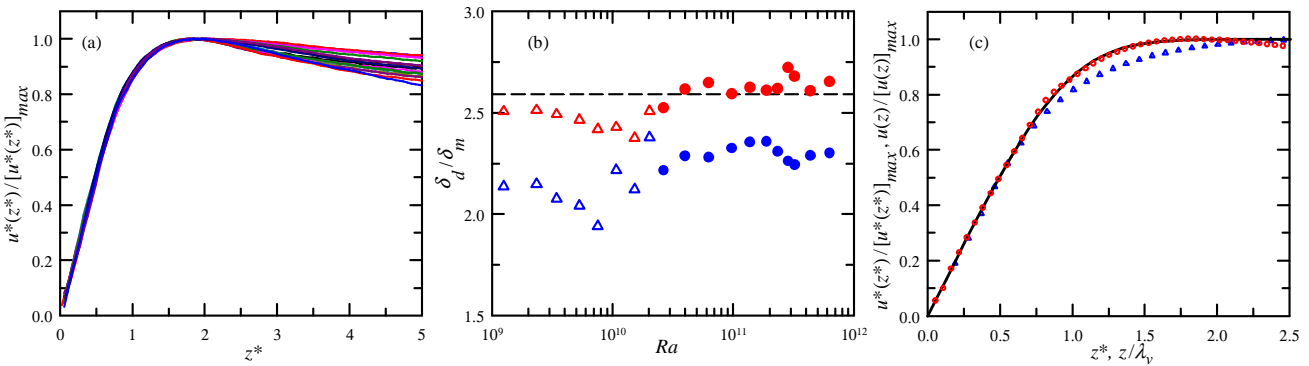}}
\caption{\label{fig:fig3} (color online) (a) Normalized profiles of $u^*(z^*)$ for all measured Ra ranging from $1.25\times10^9$ to $6.4\times10^{11}$. (b) The shape factors $\delta_d/\delta_m$ of $u^*(z^*)$ (red symbols) and $u(z)$ (blue symbols) vs. Ra for the small-cell (open triangles) and large-cell (solid circles) data. The dashed line represents the value of 2.59 for the Prandtl-Blasius BL. (c) Comparison among several velocity profiles: $u^*(z^*)$ (red circles), $u(z)$ (blue triangles), and the Prandtl-Blasius profile (black line). The data were measured at Ra $=1.9\times10^{11}$.}
\end{center}
\end{figure*}

Figure \ref{fig:fig2} shows a 200-second time trace of
$\delta_v(t)$, normalized by the most-probable thickness
$\delta_v^{mp}$. One sees that the fluctuating $\delta_v(t)$
exhibits intermittent features, i.e., the amplitude of
$\delta_v(t)$ can be much larger than its mean or most-probable
value. We find that, for all measured values of Ra,
$\delta_v^{mp}$ coincides with the viscous BL thickness
$\lambda_v$ (to within $4\%$) \cite{sun08jfm}, defined through the
slope of the time-averaged velocity profile at the plates that
have been previously studied extensively. The probability density
functions (PDF) of $\log_{10}(\delta_v/\delta_v^{mp})$ and
$\delta_v/\delta_v^{mp}$ are plotted in Figs. \ref{fig:fig2}(b)
and \ref{fig:fig2}(c) respectively. For either PDF there is
reasonable collapse for all measured Ra from both cells, which
suggests a universality of the BL dynamics in turbulent RB system.
In addition, it is seen that the tails of the PDFs at large
$\delta_v$ ($> \delta_v^{mp}$) can be described by a decaying
exponential distribution, whereas those at small $\delta_v$
($\lesssim \delta_v^{mp}$) satisfy a lognormal statistics. The
mechanism that leads to the different distributions of $\delta_v$
below and above its most probable values is at present unknown and
would be a challenge for future theoretical studies. However, it
is clear that measurement made at a fixed position in the
laboratory frame and near the boundary layer will sample a mixed
statistic (one pertaining to the boundary layer and the other pertaining to the bulk), because that position will be sometimes inside and
sometimes outside of the fluctuating BL. To achieve a
clean separation of the two types of dynamics in the measured
velocity, we define the time-dependent relative vertical position
$z^*(t)$ with respect to $\delta_v(t)$, i.e.,
\begin{equation}
z^*(t)=z/\delta_v(t),
\end{equation}
and hence $z^*$ is the rescaled distance from the plate in unit of
BL thickness. The mean velocity profile $u^*(z^*)$ in this
time-dependent frame is then calculated by averaging all values of
$u(z,t)$ that were measured at different discrete time $t$ but at
the same relative position $z^*$, i.e.,
\begin{equation}
u^*(z^*)=\langle u(z,t)|z = z^*\delta_v(t)\rangle.
\end{equation}
The results are shown in Fig. \ref{fig:fig3}(a), normalized by the maximum
value of each profile $[u^*(z^*)]_{max}$. One sees excellent
collapse of all profiles for the regions inside and around the
viscous BL ($z^*\lesssim2$), suggesting a universal BL profile for
all Ra measured in the two cells. For comparison, we also
calculate the time-averaged velocity $u(z)=\langle u(z,t)\rangle$ for each fixed position $z$. We note that
$u^*(z^*)$ is more universal than $u(z)$ in the sense that
profiles for different values of Ra collapse better in the
time-dependent frame than they do in the laboratory frame.

To characterize the shape of the $u^*(z^*)$ profile more
quantitatively, we examine its shape factor $\delta_d/\delta_m$,
where
\begin{equation}
\delta_d=\int_0^{\infty}\{1-\frac{u^*(z^*)}{[u^*(z^*)]_{max}}\}dz^*
\end{equation}
and
\begin{equation}
\delta_m=\int_0^{\infty}\{1-\frac{u^*(z^*)}{[u^*(z^*)]_{max}}\}\{\frac{u^*(z^*)}{[u^*(z^*)]_{max}}\}dz^*
\end{equation}
are the displacement thickness and the momentum thickness,
respectively \cite{schlichting04}. Because of the zero mean-flow
in the central region of a closed convection cell \cite{xia03pre},
$u^*(z)$ decays after reaching its maximum value and hence the
above integrations are evaluated only over the range from $z^*=0$
to where $u^*(z^*)=[u^*(z^*)]_{max}$. For comparison, we also show
the shape factors of the $u(z)$ profiles measured in the
laboratory frame, which are represented by blue symbols in Fig.
\ref{fig:fig3}(b). It is seen that for all values of Ra,
$\delta_d/\delta_m$ of $u(z)$ are clearly smaller than 2.59
--- the value of a Prandtl-Blasius laminar BL profile. This is because the time-averaged BL
quantities obtained in the laboratory frame are contaminated by contributions from the bulk that has a different dynamics. It is also seen that the shape factors of $u^*(z^*)$ (red symbols), measured in the time-dependent frame,
are much closer to the Prandtl-Blasius value. This suggests that the algorithm using the dynamical coordinates can effectively disentangle
the mixed dynamics inside and outside the fluctuating BLs. To see this more clearly, we
directly compare in Fig. \ref{fig:fig3}(c) the velocity profiles
$u^*(z^*)$ (circles) and $u(z)$ (triangles),  based on the same data set (Ra $=1.9\times10^{11}$) but measured in the time-dependent and
laboratory frames respectively, with the theoretical Prandtl-Blasius
profile (solid line) \cite{schlichting04}. Here, to make a proper comparison, the initial slope of the Prandtl-Blasius profile is matched to that of the measured profiles by adjusting the path length $x_0$ in the similarity parameter $\eta=z\sqrt{u_{max}/2x_0\nu}$ \cite{thess07prl}. ($\nu$ is the kinematic viscosity of water.) An excellent coincidence between $u^*(z^*)$ and the Prandtl-Blasius profile is seen for $z^*\lesssim2$. Whereas the values of $u(z)$ are clearly lower than those of the Prandtl-Blasius profile in
the region around the viscous BL ($0.7\lesssim
z/\lambda_v\lesssim2$).

\begin{figure}
\begin{center}
\resizebox{1\columnwidth}{!}{%
  \includegraphics{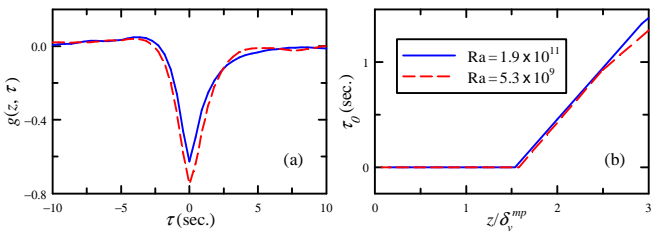}}
\caption{\label{fig:fig4}  (color online) (a) Cross-correlation function $\textrm{g}(z, \tau)$ between $\delta_v(t)$ and $u(z,t)$ measured at $z=0.76\delta_v^{mp}$. (b) The normalized peak position $\tau_0$ as a function of $z/\delta_v^{mp}$.}
\end{center}
\end{figure}

With the measured $\delta_v(t)$, we can now study some of its
dynamic properties and find out what is responsible for its
thickness fluctuation. We calculate the cross-correlation function
between $\delta_v(t)$ and $u(z,t)$, i.e., $\textrm{g}(z,
\tau)=\langle[\delta_v(t)-\langle \delta_v(t)\rangle][u(z,
t-\tau)-\langle u(z, t)\rangle]\rangle/\sigma_{\delta}\sigma_u$,
for each measuring vertical position $z$ and for each Ra. Here,
$\sigma_{\delta}=\sqrt{\langle[\delta_v(t)-\langle
\delta_v(t)\rangle]^2\rangle}$, $\sigma_u=\sqrt{\langle[u(z,
t)-\langle u(z, t)\rangle]^2\rangle}$, and $\langle\ldots\rangle$
denotes a time average. Figure \ref{fig:fig4}(a) shows two
examples of $\textrm{g}(z, \tau)$ measured inside the viscous BL.
The presence of a strong negative peak for both correlation
functions and the fact that the peaks are located at $\tau_0=0$
suggest that the thinning of the viscous BL and the increasing of
$u(z,t)$ occur simultaneously. We further note that $\tau_0=0$
holds for $z\lesssim 1.5 \delta_v^{mp}$ for all Ra
investigated [see, e.g., Fig. \ref{fig:fig4}(b)] and hence it is a
BL property. It is further found that $\delta_v(t)$ and $u(z,t)$
are highly correlated inside and around the viscous BL for all
measured Ra [see, e.g., Fig. \ref{fig:fig4}(a)]. This
suggests that the fluctuations of horizontal velocity measured
locally close to the plate (e.g. at a single point) can be used as
an instantaneous measure of the fluctuations of the viscous BL
thickness $\delta_v(t)$, whereas the measurement of $\delta_v(t)$
itself would require measuring the entire velocity profile.

\begin{figure}
\begin{center}
\resizebox{1\columnwidth}{!}{%
  \includegraphics{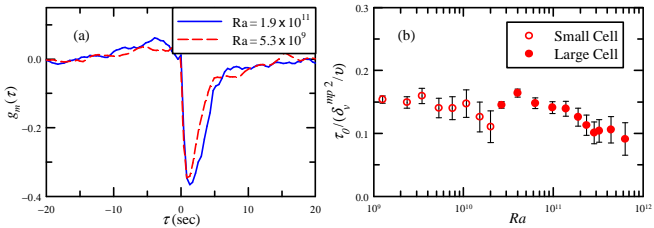}}
\caption{\label{fig:fig5} (color online) (a) Cross-correlation function $\textrm{g}_m(\tau)$ between $\delta_v(t)$ and $u_m(t)$. (b) The normalized peak position $\tau_0$ of $\textrm{g}_m(\tau)$ as a function of Ra.}
\end{center}
\end{figure}

From Fig. \ref{fig:fig4}(b) it is seen that for $z \gtrsim 1.5
\delta_v^{mp}$ a positive $\tau_0$ is obtained, suggesting that
the zero-time delay is only a boundary property. This can be made
more clearly by examining the cross-correlation function
$\textrm{g}_m(\tau)$ between $\delta_v(t)$ and $u_m(t)$, with the
instantaneous maximum velocity as a representative of the
magnitude of the mean wind. Figure \ref{fig:fig5}(a) shows
$\textrm{g}_m(\tau)$ as a function of time lag $\tau$. Two
features are worthy of note: (i) There exists a strong negative
correlation between $\delta_v$ and $u_m$, which implies that the
larger $u_m$ is, the smaller $\delta_v$ becomes; (ii) The negative
peak of $\textrm{g}_m(\tau)$ is located at a positive time lag
$\tau_0$, indicating that the fluctuations of $u_m$ leads the
variation of $\delta_v$. This is because a time delay is needed to
transfer the momentum from the mean wind to the interior of the
BL. Figure \ref{fig:fig5}(b) shows the Ra-dependence of the $\tau_0$, normalized by the typical timescale $\delta_v^{mp2}/\nu$ of
momentum transfer across the BL via viscosity. It is seen that
$\tau_0/(\delta_v^{mp2}/\nu)$ varies around 0.14 for Ra
$\lesssim10^{11}$ and decreases slightly with increasing Ra for Ra
$\gtrsim10^{11}$. Here, the decrease of
$\tau_0/(\delta_v^{mp2}/\nu)$ at high Rayleigh numbers suggests an
increased contribution from Reynolds stress to the total shear
stress inside the BL \cite{sun08jfm}. Direct study of shear stress
around the viscous BL further reveal that Reynolds stress becomes
more important than viscous stress for Ra $\gtrsim10^{11}$ and hence the BL changes gradually from being laminar to a more turbulent one.

Finally, we check the scaling properties of $\delta_v(t)$. We use
$\delta_v^{mp}$ as the typical viscous BL thickness. The best
power-law fits give $\delta_v^{mp}/H\sim0.64\text{Re}^{-0.49\pm0.02}$ for the small
cell and $\delta_v^{mp}/H\sim0.60\text{Re}^{-0.51\pm0.02}$ for the large cell, where Re $=[u^*(z^*)]_{max}H/\nu$ is
the large-scale Reynolds number. All these results are in line
with those obtained using the time-averaged profiles measured in
the laboratory frame \cite{sun08jfm} and the measured
$\delta_v^{mp}/H$-Re scaling exponents are in good agreement with
the theoretical value of $-0.5$ for the Prandtl-Blasius BL, further
confirming the conclusion of the Prandtl-Blasius-type laminar BL in RB
system \cite{gl, sun08jfm}.

We thank Chao Sun for providing us the PIV data measured in the
small cell. This work was supported by the Research Grants Council
of Hong Kong SAR (Nos. CUHK403806 and 403807). Q. Z. thanks
supports of Shanghai NSF (No. 09ZR1411200), Chenguang project
(No. 09CG41), and RFDP of Ministry of Education of China (No. 20093108120007).

\end{document}